\documentclass[12pt,a4paper]{article}
\usepackage {amsmath,amsthm,mathrsfs}
\usepackage{geometry}
\usepackage{titlesec}
\usepackage{sectsty}
\usepackage{natbib}
\usepackage{multirow}
\usepackage[hidelinks]{hyperref}
\sectionfont{\fontsize{14}{15}\selectfont}
\subsectionfont{\fontsize{13}{15}\selectfont}

\titleformat*{\section}{\LARGE\bfseries}
\titleformat*{\subsection}{\Large\bfseries}

\usepackage{caption}

\usepackage[dvipsnames]{xcolor}

\usepackage[pdftex]{graphicx}
\usepackage{booktabs}
\usepackage{float}
\usepackage{setspace}
\usepackage{placeins}
\usepackage{subcaption}
\DeclareMathOperator{\plim}{plim}

\title{\Large Peer Prediction for Peer Review: Designing a Marketplace for Ideas\\}
\author{\large Alexander Ugarov\footnote{I appreciate thoughtful suggestions provided by Arya Gaduh (University of Arkansas), Grant Schoenebeck (University of Michigan), Nihar Shah (CMU), Andy Brownback (University of Arkansas), and the seminar participants at the University of Arkansas and the University of Oklahoma. All the mistakes and shortcomings are my own.}}

\begin{document}
\maketitle
\onehalfspacing
\begin{abstract}{The paper describes a potential platform to facilitate academic peer review with emphasis on early-stage research. This platform aims to make peer review more accurate and timely by rewarding reviewers on the basis of peer prediction algorithms. The algorithm uses a variation of Peer Truth Serum for Crowdsourcing \citep{radanovic_incentives_2016} with human raters competing against a machine learning benchmark. We explain how our approach addresses two large productive inefficiencies in science: mismatch between research questions and publication bias. Better peer review for early research creates additional incentives for sharing it, which simplifies matching ideas to teams and makes negative results and p-hacking more visible.}
\end{abstract}

\newpage
\section{Inefficiencies in Science}

One of the most groundbreaking discoveries of 20th century came to life through accidental communication of two independent research groups. In their famous Nature paper (\citet{watson_molecular_1953}), Watson and Crick  acknowledged that the DNA model was inspired by unpublished experimental results of M. Wilkins and R. Franklin.  They received these results in at least two ways: through a conference presentation and through a report of R. Franklin to her funding organization. Watson and Crick worked at the Cavendish laboratory in Cambridge, while Wilkins and Franklin worked in the King's College.  Both groups eventually ended up with Nobel prizes, but this success unnecessarily relied on luck.  

Economists often view innovations as something which comes from combining different existing ideas \citep*{weitzman_recombinant_1998, berliant_knowledge_2008, lucas_ideas_2009, jr_knowledge_2014}. Due to increasing specialization in science these ideas more and more often come from different people \citep{jones_as_2011}. Suppose that one researcher has a unique research question (idea 1) and another scientist knows a method suitable for solving this question (idea 2). If these scientists connect and share these ideas, the potential research project happens. If there is no connection between the two scientists, the potential project using these ideas becomes much less likely. This inefficiency slows down the progress of science because it prevent the formation of novel ideas and hence reduces the average quality of active projects. 

Consistent with that view of innovation as recombination of ideas, recent empirical studies show that physical co-location and personal connections have a significant effect on research productivity \citep{azoulay_toward_2018, akcigit_dancing_2018, jaravel_team-specific_2018}. This suggests that the exchange of knowledge and ideas happens largely through informal networks of department colleagues, friends and collaborators. Other researchers access the research output only in the form of finished publication or conference presentation \citep*{partha_toward_1994, thursby_prepublication_2018}. This limitation reduces the pool of potential new ideas and research projects and should slow down the progress of science.

Existing empirical research tends to indicate that greater disclosure of intermediate results facilitates innovation. \citet{murray_mice_2016} demonstrate that open access to certain types of genetically modified mice led to higher research output and higher diversity of questions and methods. The field experiment conducted on TopCoder platform by \citet{boudreau_open_2015} shows that coders produce better solutions in the open collaboration environment when they can share their code in the process as compared to the environment in which only final solutions are shared. 

The practice of scientists disclosing only final results also leads to publication bias in the form of under-reporting of negative results \citep*{song_extent_2009,franco_publication_2014}. Scientists often decide to not publish their null results even if their methodology is sound. It leads to overestimation of treatment effects in empirical studies and slower refutal of existing theories. The commonly proposed solution is to pre-register all the experimental studies and to catalogue their findings \citep{nosek_preregistration_2018}. This motivation stands behind the creation of AllTrials project which aims to collect information on all the clinical trials conducted regardless of their results. The US federal project ClinicalTrials.gov (run by United States National Library of Medicine (NLM)) similarly collects the data on clinical trials conducted in the US. 

The sharing of early research results in science is currently limited and hence there is large potential for more research openness. \citet{thursby_prepublication_2018} find that, depending on the field, between 45\% to 75\% of scientists share their drafts before journal submissions, but less than 15\% do so on the conceptual stage. These limitations on sharing are partially due to tradition and partially due to technological deficiencies we hope to address. Many scientists see several downsides of early sharing which include scooping and potential negative reputation impacts from disclosing unverified results. The benefits of early sharing are less visible with current technology. Existing open science platforms do not provide feedback to early research and sharing early research does not contribute to researcher's reputation and visibility outside of their close networks. Our platform aims to solve these technological issues by creating incentives for sharing in form of research feedback and reputation gains. 

More open sharing of research questions and interim results should help to solve the two inefficiencies we identify in academic practice: limitations in matching of research ideas, and the publication bias. Publishing questions and early research results would help to combine ideas and resources across research groups and geographical boundaries. Interim results similar to pre-registration of studies would increase knowledge about negative results. Finally, better knowledge of competing projects would also reduce privately suboptimal redundant research.

\vspace{30pt}

\section{Platform's Overview}
The project aims to create an open science repository with an incentive-compatible peer-review mechanism\footnote{Simply speaking, the mechanism is incentive-compatible if agents get higher payoffs by acting truthfully. For example, ordering in a restaurant is incentive-compatible because a restaurant's client would get the highest expected payoff by truthfully stating their preferred choice. This definition encompasses strategy-proof mechanisms and Bayesian Nash equilibrium incentive-compatible. In strategy-proof mechanisms agents always receive higher or equal payoffs by reporting truthfully regardless of choices of others. A Bayesian Nash equilibrium incentive-compatible mechanism has at least one equilibrium in which acting truthfully gives the best expected payoff conditional on others acting truthfully.} The platform would have an interactive database of research projects at all stages of work from a research idea to a final paper with several key features:
\begin{enumerate}
\item Publishing research projects on stages of idea/proposal/working paper.
\item Requesting and publishing referee reports (including reports anonymous for the author and/or third parties).
\item Publishing comments.
\item Rating research projects and referee reports.
\item Hosting grant competitions.
\end{enumerate}

The platform\footnote{With its beta-version now being published at \href{https:\\hivereview.org}{Hivereview.org}} aims to reward sharing of interim research results and questions in two ways. First, it will give researchers access to early peer-review and pools of potential collaborators. Second, it would increase reputation and visibility of users sharing valuable questions, interim results, data sets, techniques and comments.

The platform should provide sufficient participation incentives for scientists at different career stages. Highly established scientists can use it to increase their footprint in science and advertise their research groups. Early career scientists would benefit by learning through the feedback and by increasing their professional exposure. 

There are several other open science repositories, including ResearchGate, Academia.edu, SSRN, PubPeer, Research Hub, and arXiv. While these platforms mostly function as preprint repositories, they do not provide formal peer-reviews with the exception of PubPeer and Research Hub. All of these platforms are not designed to provide accurate feedback on early research. They do not host research ideas and so provide limited opportunities to direct research and seek collaborators. For these reasons, we believe that a different platform is still needed.

\vspace{30pt}
\section{Previous Research}

The idea of separating dissemination of papers with their peer review is not new. When the Internet made online access to published papers much easier, it reduced one important rational for using peer review as a necessary condition for publishing, because the costs of Internet publishing are much lower.  One of the earliest proposals for creating a platform to publish both papers and their reviews was made by Yann LeCun\footnote{The proposal is published \href{https://yann.lecun.com/ex/pamphlets/publishing-models.html}{here}, but requires a password at the moment of writing this paper.}.  \citet{soergel_open_2013} discussed different approaches to transparency of referee reports and proposed an open academic review platform which makes both papers and their reviews open to users. The platform \href{https://openreview.net/}{Openreview.net} based on this proposal is currently used for reviewing conference submissions in several large computer science conferences including the International Conference on Machine Learning (ICML).

The novelty of our approach is in additional incentives for accurate peer review which can stimulate more intermediate sharing. There are well-known examples of dishonest behaviour emerging when peer review starts affecting reviewers personal outcomes such as in a conference program selection \citep{shah_challenges_2022}. Hence it is crucial that any proposed mechanism will be incentive-compatible in the sense that self-interest does not drive actors to dishonest actions.  \citet{srinivasan_auctions_2021} describe an alternative to ours incentive-compatible mechanism which can be used for selecting conference or journal submissions. The proposed mechanism involves authors of paper submissions bidding for reviewing slots in a VCG auction and then using the auction proceeds to reward reviewers based on quality of their submissions. The quality evaluations is done by other mechanism participants based on a version of a peer prediction algorithm developed by \citet{kong_information_2019}. Given a different goal of our proposal, we decided to opt for a simpler version of the mechanism which does not require common prior from users and provides additional defence against non-truthful equilibria.

\vspace{30pt}
\section{Reputation Metric}
As some other open science platforms, the project would use reputation metrics to reward users for services useful to the community such as sharing open questions and giving accurate feedback. In contrast to existing platforms however, we specifically design the reputation metric mechanism to reward for truthful feedback to make the reputation an informative signal of researcher's quality. 

The public reputation metric determines the visibility of users’ projects and impacts of their peer evaluations. The platform will have an incentive-compatible peer review system for two types of user's submissions: research projects and referee reports. Each user in the system would have an ability to rate all the projects and all the referee reports of other users\footnote{Subject to potential limitations on the total impact or number of ratings to limit the power of overactive users.}. Users will be able also to like comments provided to projects and referee reports, but likes will not affect anyone's reputation.

We expect the reputation metric to proxy for the quality of the user's research output. The reputation of user $i$ is a sum of their initial reputation $R_i(0)$ (such as academic rank and H-index), reputation of the user's projects $R^p_i(t)$ , user's referee reports $R^r_i(t)$ and accuracy score of peer evaluations $E_i(t)$ given by user $i$ for all the projects and referee reports they reviewed. The accuracy score of peer evaluations $E_i(t)$ for user $i$ is the sum $E^p_i(t)$ of the accuracy scores for all the research projects rated by $i$ plus the sum $E^r_i(t)$ of accuracy scores for all the referee reports rated by the user. We can describe the reputation by the following formula if abstracting from weighting of different components:

$$R_i(t)=R_i(0)+R^p_i(t)+R^r_i(t)+E_i(t), E_i(t)=E^p_i(t)+E^r_i(t)$$

Why would users value their reputations? The reputation would be valuable if it becomes an informative signal of the user's quality as a researcher and can affect their visibility and their careers\footnote{Users would not be able to observe the composition of others' reputations. If users decide that the reputation of research projects is much more informative than the rest of the metric, they would lose incentives for accurate peer evaluations. This would, in turn, erode the quality of peer evaluations and can unravel the rest of the reputation metric.}. The reputation metric is informative and valuable if it improves the expectation of future success conditional on other publicly available information\footnote{The signal is still informative even if its predictive power comes solely from self-fulfilling expectations.}: 

An informative signal would still have zero value if its users do not know that it is informative. The sufficient condition is to require that it is common knowledge that the ratings affecting the reputation are (largely) truthful. This makes it extremely important to rely on incentive-compatible accuracy ratings.

Several successful platforms rely on reputation metrics to reward users' contributions. StackOverflow calculates reputation based on votes given to users' questions and answers. This reputation gives some minor privileges in accessing website features. Quora is one the most well-known question and answer platforms and it determines visibility of users' answers based on average views and upvotes of their previous answers. Code repositories on Github receive ``stars'' from other users and the number of these stars signals the value of the repository and proximally the ability of their creators. Besides reputation, many users of online collaboration or crowdsourcing platforms can have intrinsic (altruistic, self-realization) motivations, which should also apply to this project. For example, \citet{chandler_breaking_2013} find that subjects prefer participating in tasks framed as more meaningful.

\vspace{30pt}

\section{Rewarding Accurate Scores}

Accuracy scores reward users for truthful ratings of research projects\footnote{Research projects include research ideas, working papers and published papers.} and referee reports. Users providing more accurate ratings expect higher scores and hence higher total reputations. In contrast, providing distorted or noisy ratings harms the accuracy score and the reputation.

The scoring algorithm for both research projects and for referee reports relies on the approach of \citet{radanovic_incentives_2016} called Robust Peer Truth Serum for Crowdsourcing (RPTSC). \rm The approach elicits hidden correlated information from multiple agents in crowd-sourcing applications, when getting this information requires some effort and no verification is available to the center. 

Each rater submits a categorical report stating their evaluation of an item. For example, a rater $w$ can be asked to categorize a third party's referee report as ''unsatisfactory'' ($u$), "satisfactory" ($s$) or "exceptional" ($e$). We need at least two raters for each report to calculate the accuracy score, because it depends on the match between the rater's $i$ rating and the rating of some randomly chosen other rater (peer) $p$. However, the mechanism is incentive-compatible even if there are no other raters as long as raters believe that other raters will eventually participate with some positive probability. 

The algorithm works as follows for any referee report $k$ if a rater $w$ gives a rating $r_{wk}$:
\begin{enumerate}
\item Randomly select $n-1$ other referee reports.
\item Sample $n$ ratings from $n$ different reports, including report $k$ but excluding the rating of rater $w$ being evaluated. 
\item If there are no other ratings for report $k$, the accuracy score is zero. If there is another rating, call it a peer rating $r_{pk}$.
\item Calculate the proportion of ratings $F(r_{wk})$ in this sample which has the same rating value $r_{wk}$.
\item The accuracy score is calculated as:
$$e_{wk}=\alpha(\tau(r_{wk},r_{pk})-1)$$
$$\tau(r_{ik},r_{pk})=\begin{cases} {1 \over F(r_{wk})}, r_{wk}=r_{pk}\\0, r_{wk}\neq r_{pk} \end{cases} $$
\end{enumerate}

Here the parameter $\alpha>0$ scales the importance of accuracy scores for the total reputation metric. We can adjust this parameter to create sufficient incentives to participate in grading. The original mechanism treats scoring as one-time event, but we treat it as an infinite horizon scoring with scoring being redone perpetually. All the original assumptions needed for the proof still hold if we extend evaluation to infinite horizon.

\citet{radanovic_incentives_2016} show that this mechanism is incentive-compatible under some plausible assumptions on beliefs\footnote{Let $P_w(x)$ denote a prior belief of worker $w$ that the answer is $x$ and $P_{p|w}(y|x)$ - posterior belief of worker $w$ that the random peer $p$ gets an answer $y$ conditional on worker $w$ getting answer $x$. Then we can write the assumptions on beliefs as follows: 1) tasks are statistically independent: $P_{p|w}(x|y)=P_p(x),\forall p\neq w$ 2) tasks are allocated randomly $P_p(x)=P_q(x), \forall p,q,x$ 3) beliefs are fully mixed $P_p(x)>0, P_{p|q}(x|y)>0$ 4) self-predicting condition: $P_{p|w}(x|x)>P_{w|p}(x|\bar x)$.}: telling the truth gives the highest expected accuracy score if others also tell the truth. Intuitively, incentive-compatibility follows, because the accuracy score rewards ratings $x$ which are surprisingly common relative to the baseline $F(x)$. More specifically, truth-telling is a subjective equilibrium of this mechanism meaning that it gives the highest payoff for any admissible belief about priors of other users. The mechanism is also individually rational: users would benefit from giving truthful evaluations, because in a truthful equilibrium the expected accuracy score is strictly positive. 

\citet{radanovic_incentives_2016} find that the RPTSC approach performs well in a peer grading field experiment. In the experiment, students enrolled in a large undergraduate class rate coding assignments of their peers by using a solution provided by TAs. There are three randomized  groups with different incentive schemes. Depending on a group, students receive bonus points for grading accuracy either as a flat payment, as a peer consistency payment or as a RPTSC payment. They find that the group rewarded according to the RPTSC mechanism produces more accurate grading reports and the difference is statistically significant at 5\% confidence level.

\vspace{10pt}
\bf Addressing Uninformative Equilibria. \rm \textit{Uninformative equilibrium} is when mechanism participants give correlated ratings without reviewing the item. For example, if everyone reviews all the referee reports as "exceptional" regardless of their quality than the ratings are perfectly matched. This strategy profile is an equilibrium of the RPTSC mechanism but it does not produce any new information about the reports being evaluated. \citet{gao_trick_2014} experimentally show that users rewarded according to several most standard peer prediction algorithms such as \citep*{miller_eliciting_2005, jurca_collusion-resistant_2007} tend to end up in uninformative equilibria.

The original RPTSC algorithm already provides partial protection against uninformative equilibria. The strategy of rating all the items identically in order to reduce cognitive effort obtains a non-positive expected payoff (accuracy scores) regardless of strategies chosen by others. Even if everyone reports the same rating, the construction of the reward function guarantees that the expected payoff is zero. Hence, raters have no incentive to make uniform reports even if it happens to be an equilibrium conditional on participation. 

However, the original RPTSC mechanism can have uninformative equilibria with positive expected payoffs whenever there is some easily available information about items. For example, raters can give uniform high ratings to all the projects containing more than 20 pages. This is an uninformative equilibrium with positive expected payoffs. Uninformative equilibrium can be particularly harmful if users decide to discriminate against some group or research topic out of bias, prejudice or animosity.

In order to deal with more sophisticated uninformative equilibria, we suggest using random forest predictions based on public descriptors $D$ as benchmarks for human prediction. We need to assume that there exists the lower bound on the frequency of rating $x$ ($P(x)>\epsilon,\forall x$). Then we estimate the distribution of the report's quality conditional on easily observable characteristics (such as word count) $Q(x, D)$ and then use $\tilde Q(x, D)=\max(\epsilon,Q(x,D))$ instead of empirical frequencies $F(x)$. This adjustment accounts for observable heterogeneity in ratings' distribution and adds regularization to avoid extremely low values. We can consider the RPTSC original mechanism as a special case of our approach. 

Let's define uninformative equilibrium as a pure strategy equilibrium with strategies mapping item's descriptors $D$ to ratings. Item's descriptors include its easily observable characteristics, such as its title, word count, number of citations and its authors' reputations. The usefulness of our augmented mechanism relies on the following two statements:

\textit{Conjecture 1: If the self-predicting condition holds for any partition based on $D$\footnote{For example, if the set of descriptors $D$ includes authors' genders and research fields it means that the condition holds both in each gender and in any set defined through the combination of gender and field.} then truthful reporting is a subjective equilibrium of the augmented mechanism.}

\textbf{Argument}: The original argument applies after redefining the set of potential items (reports) appropriately. The problem is to show that there are non-trivial environments for which these conditions hold.

\textit{Conjecture 2: Expected payoffs of any uninformative equilibrium in the augmented mechanism converge to zero with N of ratings.}

\begin{proof}: Let $r:D\to M$ be an equilibrium strategy profile resulting in a ratings' distribution $P(x,D)$. Moreover, as reported ratings do not depend on signals, then $P(x|y;D)=P(x,D)$. We know from \citep{malley_probability_2012, kruppa_probability_2014} that the random forest algorithm generates consistent estimates of probabilities, meaning that $\plim_{N\to\infty}\tilde Q_N(x,D)=P(x,D)$ (regularization doesn't affect consistency). Hence the expected accuracy score converges to:
$$\plim_{N\to\infty}E[\alpha(\tau(r_{wk},r_{pk})-1)]=\alpha \left[P(r_{pk}|r_{wk}) \lim_{N\to\infty}E\left({1\over \tilde Q_N(x,D)}\right)-1\right]=$$
$$\alpha \left[P(r_{pk}|r_{wk}) \plim_{N\to\infty}\left({1\over  \tilde Q_N(x,D)}\right)-1\right]=\alpha \left[{P(x,D)\over P(x,D)}-1\right]=0$$
Here the first equality follows from independence of tasks implying the independence of the estimated proportion $\tilde Q(x,D)$ and the reports for task $k$,  the second equality follows because rv's ${1\over \tilde Q_N(x,D)}<{1\over \epsilon}$ are uniformly bounded and hence uniformly integrable, and the third equation follows by the Slutsky's theorem.\end{proof}

In the augmented mechanism, the random forest algorithm provides a benchmark against which human raters have to compete. Human raters receive positive payoffs only by outperforming machine learning. it is possible for human raters to do, because they have access to semantic information not available to the algorithm. This makes our approach similar to adversarial machine learning algorithms except our approach relies on human raters as adversaries.

\vspace{10pt}
\bf Scoring Research Projects. \rm We ask users two questions to evaluate any research project:
\begin{enumerate}
\item Potential contribution:  by how much the results of the project would shift our beliefs on important questions?
\item Research design: does the chosen methodology provide a confident answer?
\end{enumerate}
A rater answers each question by choosing one of several potential answers. For example, the research question's importance can be rated as "Unclear", "Minor contribution", "Important for the research field", "Important for the discipline". 

The scoring starts immediately after the project's upload or significant update and continues until the necessary numbers of ratings is received from the users. Users cannot observe ratings of others or the aggregate rating of the project until the scoring is complete. This limitation aims to prevent users from using the current rating as the focal point for collusion.

The platform assigns numeric scores to all the ratings provided according to the PTSC mechanism above as soon as the project has a minimum number of ratings to calculate the scores. The scores and hence their impact on rater's reputation are constantly updated as new raters arrive. For example, the project can be rated highly by the first two raters and low by the third raters with negative initial impact on the third rater's reputation. However, the third rater's accuracy score and reputation would go up if subsequent raters also rate the project low. 

Updates significant enough to affect answers to the evaluation questions necessitate resetting accuracy scores. Resetting is done by authors' request separately for each of the question. For example, the authors can request resetting the methodology ratings to reflect significant methodological improvements.

\vspace{10pt}
\bf Referee Reports. \rm Any referee report becomes available for scoring by other users immediately after its submission. Any user, including the authors of the refereed project, can rate a referee report $k$ either as ''unsatisfactory'' ($u$), "satisfactory" ($s$) or "exceptional" ($e$).\footnote{The mechanism is incentive-compatible for any finite number of labels, but keeping the number of labels low would help to keep it transparent for users.}. The scoring becomes final after some period of time since publishing or after the sufficient number of reviews is provided. The referee report's ratings are not available to any users until the scoring is finished, but becomes visible to everyone afterwards.

\vspace{10pt}
\bf Reputation Updates. \rm As a reminder, the user's $i$ reputation is the sum of several components, which include the reputation of research projects $R^p_i$, the reputation of user's referee reports $R^p_i$,  and the accuracy scores of submitted ratings for both projects $E^p_i$ and reports $E^r_i$. Each reputation component is a sum of ratings of corresponding individual items. For example, the total rating of the referee report $s$ is the sum of ratings $x_{js}$ given by users weighted over different categorical levels $k$:

$$r^r_s=\sum_s \sum_k \alpha_k \sum_j I(x_{js}=k)$$

Then the total contribution from the user's referee report to their reputation is just a simple sum of ratings of their scored referee reports $O^r_i$:
$R^r_i=\sum_{s\in O^r_i} r_s$. The total rating of any research project and the reputation component of research projects is calculated in a similar way with some possible exception of some scaling constant.

The accuracy components of the reputation $E^r_i,E^p_i$ are the sums of accuracy scores of individual items rated by the user and weighted by some positive constants $\beta^r, \beta^p$. Let $\Omega^r_i$ denote the set of referee reports rated by user $i$ and $\Omega^p_i$ is the set of projects rated. Then the accuracy components of the reputation are:
$$E^r_i=\beta^r \sum_{s \in \Omega ^r_i}e_{is},\hbox{ }E^p_i=\beta^p \sum_{s \in \Omega ^p_i}e_{is}$$

Note that the accuracy score is going to change over time as more and more users provide evaluations. If the equilibrium stays truthful, it results in lowering the expected score's variance and making it a more precise measure of quality. This dynamic aspect also makes the mechanism more robust to untruthful equilibria. If a user expects that the equilibrium is going to be untruthful, the expected user's score from submitting an untruthful report will be zero. However, if a user expects that the truthful equilibrium would emerge in the future with some probability then reporting truthfully would maximize their expected score.

\vspace{10pt}

\bf Alternative Scoring for Research Projects. \rm In order to eliminate uninformative equilibria, the platform can instead use the feedback provided through the publication process. It can then calculate the scores based the quadratic score function \citep{selten_axiomatic_1998}. Quadratic score function is a proper score function mapping on the reported probabilistic belief about an event and the event's actual realization to one number. By definition of the proper scoring function, its expected value is the highest when the true probability is used. Hence the mechanism is incentive-compatible in the strictest sense as long as users are risk-neutral (care only about the expected value).

In order to get a project rated, the author posting it on the platform has to specify one or more publicly verifiable binary measures of the project's success. As an example, the author can ask if paper(s) resulting from the project would be cited at least 10 times within three years after the project's progression to the final stage. The research project's rating is an estimate of the probability for that event. If user $i$ predicts that the probability of this event is $p_{ik}$ then the accuracy score of this rating will be:
$$e_{ik}=S(p_{ik},o_k)-S(p_{Bk},o_k))$$
$$S(p,o_k)=1-2(o_k-p)^2$$
Where $o_k$ is the indicator of the event:$o_k=1$ if it occurs (e.g., the papers get cited at least 10 times), $o_k=0$ otherwise. The component $S(p_{Bk},o_k)=1-2(o_k-p_{Bk})^2$ represents the benchmark score to re-scale the accuracy. The value $p_{Bk}$ is the benchmark prediction we obtain by using either the average community prediction or the algorithmic prediction. Algorithmic prediction would use basic machine learning techniques to estimate the probability based on verifiable project's characteristics such as completion stage, word count, and reputation of its authors. It is crucial that the benchmark score does not depend on user $i$'s actions and hence does not change their optimal strategy conditional on participation.

If the benchmark score comes from the AI, users improve their reputation by predicting better than the algorithm.  In contrast to the community average prediction, AI rating can be made public almost immediately and before any private ratings. Users will be able to evaluate potential reputation gains by comparing their private beliefs against the AI rating. It should motivate users to seek underrated/overrated projects and incorporate new private information into their predictions. The simplest AI prediction involves just a constant score for all the items within the category (projects/reports).

Quadratic score function is an incentive-compatible mechanism to elicit beliefs under two plausible assumptions  \citep{selten_axiomatic_1998}. First, we need to assume that users prefers more to less reputation. Second, we need to assume that users are risk-neutral or, in other words, they care only about their expected reputation and not about other moments of its potential distribution. As we argue before, the reputation should be valuable for users if it is an informative signal of their quality as researchers and if users value how others perceive their quality\footnote{This assumption may not hold for people expecting to leave academic community in near future (retirement, death, grave ethical violations).}. There is no reason to know in advance if the users are going to be risk-averse to reputation, but the bias from risk-averse users is unlikely to make reported ratings uninformative. Both theory and laboratory experiments show that risk-averse raters bias their reported beliefs by over-reporting low beliefs and under-reporting high beliefs \citep*{armantier_eliciting_2013, andersen_estimating_2014}. This bias does not change the strictly monotonic relationship between beliefs and reports and can be ex-post corrected \citep{offerman_truth_2009}. 

Several existing platforms already use proper scoring rules or other reputation-based prediction markets to measure beliefs of their users. Metaculus allows users to bet their reputation on outcomes of future events by using the combination of the log-score and other proper scoring functions. On 30 March 2021 Metaculus had almost 16 thousand registered users (forecasters)\footnote{Based on its API: https://www.metaculus.com/api2/users/}. Their internal report show that median reported beliefs (community beliefs) from the start at 2016 to February 2021 had a Brier proper score of 0.122 meaning significant gain over the random reporting\footnote{Unfortunately, there is no published comparison of forecasts between prediction markets with real money and Metaculus.}. Hollywood Stock Exchange platform existing since 1996 has several surprisingly accurate predictions of box-opening takes for movies \citep{mann_power_2016}.

\vspace{30pt}
\section{Potential Markets of Academic Services}

While a widely-recognized reputation metric (similar to citation counts) can create strong incentives for sharing research and giving feedback, it might have no effect on motivation of participants in the project's early stages. For this reason, we are implementing an parallel market-based mechanism in which providing academic services makes easier to benefit from services provided by others.

We consider creating three markets of academic services with platform-specific tokens used across all the markets. The first market allows for exchanging research output including open research proposals, questions and results. The market of reviewing allows trading refereeing services. Finally, the market for other services gives access to additional platform features requiring intensive computation. 

\begin{table}[htbp] \centering
\begin{tabular}{l|c|c|}
\multicolumn{3}{l}{\textbf{Market of Academic Services}}\\
\hline \hline
\bf Services & \bf Earn tokens & \bf Spend tokens\\
\hline
\multirow{2}{*}{Ideas} & -Publishing open research ideas & -Adopting open ideas\\
 & -Answering open questions & -Bounties on research questions\\
\hline
\multirow{2}{*}{Reviewing} & -Writing (good) referee reports & -Requesting referee reports \\
 & -Rating projects and reports & - \\
\hline
Other services & - & -AI assistance, IFPS storage(?) \\
\hline
\end{tabular}
\end{table}

On the market of referee reports, authors can request a peer review by putting a bid on their project. A reviewer providing a referee report of satisfactory quality receives tokens in the amount of the bid. It accordingly reduces the stock of tokens owned by the authors (buyers). The total number of compensated referee reports can be determined by the project's authors. For example, authors can limit the number of needed reviews to three to satisfy the requirements of a specific conference.
 
The market of ideas trades two types of services. First, both researchers and potential donors can direct research by putting bounties on research questions. Researchers claim the bounty once their papers or research concepts successfully addresses the question. This market allows open grant applications and simplifies access to researchers for smaller donors who cannot afford hiring their own scientists to evaluate grant proposals. Second, researchers can post open research ideas and proposals which, if adopted, also increase their stock of tokens while reducing the stock of adopters.

Finally, users earn tokens when they provide accurate and informative ratings of research projects and referee reports. These rewards use an incentive-compatible mechanism in which reporting truthfully is an equilibrium. In contrast, highly inaccurate or uninformative ratings may reduce the user's stock of tokens. The calculation of scoring rewards is the same as the one used for the reputation updates.

The supply of tokens is linked to the number of registered users to reduce variation in tokens' value. The platform issues new tokens when a new user joins the platform. To make peer review incentives more salient and to simplify funding from donors, the platform can also make tokens convertible into USD.

\vspace{30pt}
\section{Conclusion}

The paper explains why improving sharing of intermediate research results should speed up scientific progress.  It also proposes a more concrete mechanism of promoting intermediate sharing with truthful peer review. This mechanism relies on using peer prediction algorithms to calculate payoffs for reviewing with these payoffs later contributing to users' influence and ability to request further services from the platform. We also present a new variation of a peer prediction algorithm which is more suited for scientific peer review and allows to eliminate uninformative equilibria.

Despite all the potential efficiency gains of the new mechanism, the success of its implementation will ultimately depend on the ability of academic culture to absorb it. There are several cultural factors limiting sharing of intermediate results, including aversion to showing incomplete work and preferences to keep more exclusionary peer-to-peer networks. Whether the benefits of the new system outweigh these cultural barriers still remains to be seen.  But regardless of the potential adoption of a more general system,  this mechanism of eliciting truthful feedback is still applicable in more contained settings, such as academic conferences and community-based research proposal evaluations.

\newpage
\singlespacing
\small
\bibliography{PPPR_wp.bib}

\end{document}